\documentclass[serif,twocolumn]{wiley-article}
\usepackage[super,comma]{natbib}
\newcommand{\absdiv}[1]{%
	\par\addvspace{.5\baselineskip}% adjust to suit
	\noindent\textbf{#1}\quad\ignorespaces
}

\usepackage{siunitx}
\usepackage{float}
\usepackage[mode=buildnew]{standalone} % requires -shell-escape
\usepackage{tikz}
\usepackage{placeins}
\usepackage{booktabs}
\usepackage{tabularx}

\graphicspath{{Figures/}}

\papertype{Original Article}
\title{Speeding up Photoacoustic Imaging using Diffusion Models}

\abbrevs{PAM, DL, SOTA, DIP, SIRS, PSNR, GAN, DDPM, ILVR, MCG, DPS, SSIM, LPIPS}

\author[1]{Irem Loc}
\author[2,3]{Mehmet Burcin Unlu}

\affil[1]{Bogazici University Physics Department, Istanbul, Turkey}
\affil[2]{Faculty of Engineering, Ozyegin University, Istanbul, Turkey}
\affil[3]{Faculty of Aviation and Aeronautical Sciences, Ozyegin University, Istanbul, Turkey}
\corraddress{Bogazici University Physics Department, Istanbul, Turkey}
\corremail{locirem@gmail.com}

\runningauthor{Loc et al.}

\begin{document}
\nolinenumbers

\begin{frontmatter}
\maketitle

\begin{abstract}
	
	\absdiv{Background:} Photoacoustic Microscopy (PAM) integrates optical and acoustic imaging, offering enhanced penetration depth for detecting optical-absorbing components in tissues. Nonetheless, challenges arise in scanning large areas with high spatial resolution. With speed limitations imposed by laser pulse repetition rates, the potential role of computational methods is highlighted in accelerating PAM imaging.
	\absdiv{Purpose:} We are proposing a novel and highly adaptable DiffPam algorithm that utilizes diffusion models for speeding up the photoacoustic imaging process. 
	\absdiv{Method:} We leveraged a diffusion model trained exclusively on natural images, comparing its performance with an in-domain trained U-Net model using a dataset focused on PAM images of mice brain microvasculature. 
	\absdiv{Results:} Our findings indicate that DiffPam achieves comparable performance to a dedicated U-Net model, without the need for a large dataset or training a deep learning model. The study also introduces the efficacy of shortened diffusion processes for reducing computing time without compromising accuracy.
	\absdiv{Conclusion:} This study underscores the significance of DiffPam as a practical algorithm for reconstructing undersampled PAM images, particularly for researchers with limited AI expertise and computational resources.

% Please include a maximum of seven keywords
\keywords{Diffusion Models, Photoacoustic Imaging, Image Reconstruction}
\end{abstract}
\end{frontmatter}

\section{Introduction}

Photoacoustic Microscopy (PAM) imaging, also called optoacoustic imaging, is a rapidly emerging field of non-invasive medical imaging\cite{Wang2014-em}. The principle is based on detecting the ultrasonic waves generated by pressurized tissues by laser beams. As the name suggests, it combines the advantages of both optical and acoustic (ultrasonic) imaging techniques. The penetration depth is much higher than purely optical imaging modalities while it can detect the optical absorbing parts of the tissue such as hemoglobin, lipids, water, and other light-absorbing chromophores\cite{beard2011biomedical}.

Similar to other imaging techniques, there is a trade between scanned area, scanning time, and spatial resolution. Scanning large areas with high resolution increases the time complexity of the whole process. In the literature, advanced scanning techniques are developed for high-speed PAM, however, the main factor that restricts the scanning speed is the laser pulse repetition rate\cite{CHO2021100291}. Unless advanced lasers are used, the best option for speeding up PAM imaging is reducing the scanning area and reconstructing the image via computational methods.

%\begin{figure}[h!]
%	\centering
%	\includegraphics[width=.40\textwidth]{scanning_time.png}
%	\caption{There is a trade of between scanned area, scanning time and spatial resolution.}
%	\label{fig:fig1}
%\end{figure}

For the past couple of years, deep learning (DL) has been used as the state-of-the-art (SOTA) computational method for medical image reconstruction, enhancement, denoising, and super-resolution. Many recent studies have illustrated that DL methods can offer solutions for problems in PAM imaging that are otherwise unsolved.

Zhao et al. \cite{https://doi.org/10.1002/advs.202003097}, used a multi-task residual dense network to speed up optical resolution PAM with low pulse laser energy. They addressed the issues of image denoising, super-resolution, and vascular enhancement simultaneously through multi-supervised learning. Sharma et al. \cite{Sharma2020-rq} employed supervised fully dense U-Net to enhance out-of-focus acoustic resolution PAM images. DiSpirito III et al. \cite{DiSpirito2021-dl}, approached the reconstruction of undersampled PAM images, i. e. an inpainting problem. They demonstrated that using supervised fully dense U-Net they were able to reconstruct artificially downsampled PAM images. All techniques up to now require a large set of PAM images to train the neural network. On the contrary, Vu et al.\cite{VU2021100266} use the deep image prior (DIP) \cite{UlyanovVL17} technique which uses an untrained neural network as an image prior. They reconstructed the undersampled PAM images, similar to DiSpirito III et al. \cite{DiSpirito2021-dl}. Based on our knowledge, this study is the only one which does not require supervised training.

In this study, we are proposing a novel method that utilizes a pre-trained diffusion model to reconstruct the degraded PAM image. This approach leverages the effectiveness of neural networks pre-trained on natural image data while eliminating the need for an extensive corpus of PAM images and mitigating the cumbersome training processes. Notably, our methodology demonstrates a degree of adaptability, offering the freedom to select the specific degradation operation, whether it be super-resolution, denoising, or inpainting. Furthermore, our proposed framework can integrate and benefit from the existing enhancement techniques.

\section{Background}
\subsection{Inverse Problems in Image Processing}

Reconstructing the degraded PAM image can be considered as a subset of inverse problems in image processing, which are by nature ill-posed. We can select the degradation operation as lowering the resolution, masking or noise adding, corresponding super-resolution, inpainting, or denoising, respectively. In this study, we approached the problem as a modified version of single image super-resolution (SIRS) and inpainting, separately.
The term super-resolution refers to techniques to enhance the resolution of an image. Similarly, image inpainting is a set of techniques that restore missing pixels in an image. In the literature, there are various approaches for solving SIRS and inpainting, the main ones are interpolation and DL-based methods. In recent years, deep learning-based approaches have proven to be superior to interpolation methods. Early studies on super-resolution/inpainting utilizing deep learning employed PSNR-based methods. Today, many SISR/inpainting techniques take advantage of deep generative models. The SOTA generative models that are used for super-resolution/inpainting are encoder-decoder networks, generative adversarial networks (GANs), flow-based models, and denoising diffusion probabilistic models (DDPMs).

Deep generative models can be used for both conditional and unconditional image generation. Unconditional diffusion models learn the underlying distribution of natural images to generate an image from pure noise. On the other hand, conditional generation requires a condition that can be in the form of text, class label, or image. For super-resolution or inpainting applications, degraded images can be used as the condition.

Conditional generation can be achieved by using the condition directly in the training process or leveraging the power of pre-trained unconditional generative models. Both approaches demonstrated remarkable performance in inverse problems in image processing \cite{li2023diffusion}. The former requires a large corpus of high-quality images of interest, the specific design of the model, and the training process. In this study, we used the latter approach, since it is more robust and flexible.

\subsection{Denoising Diffusion Probabilistic Models}

\begin{figure}[h]
		\centering
		\includegraphics[width=.48\textwidth]{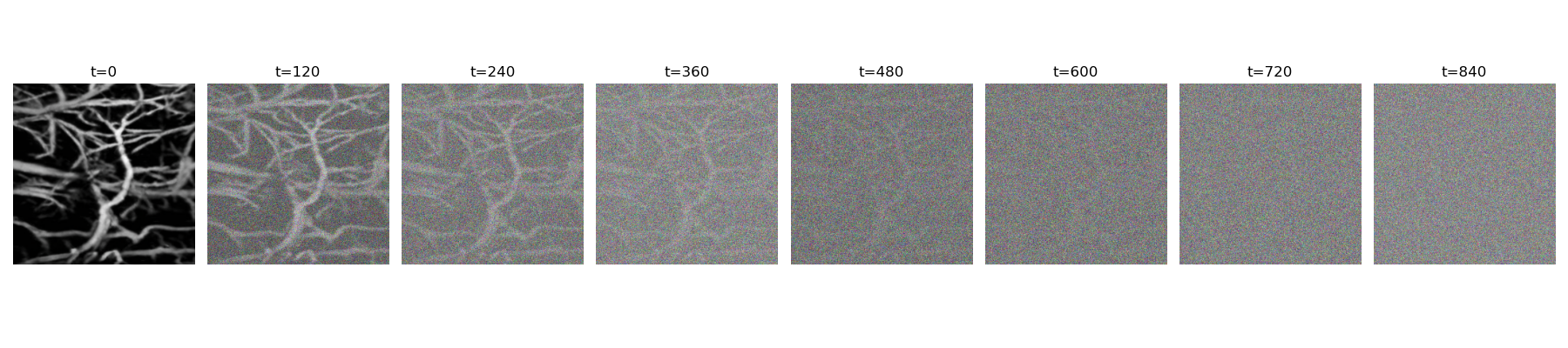}
		\includegraphics[width=.48\textwidth]{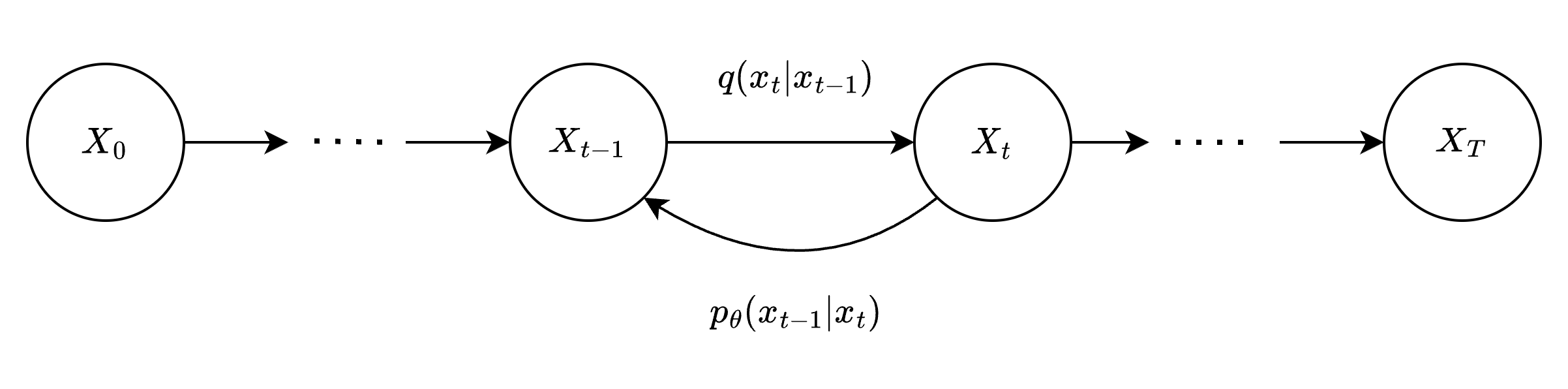}
		\caption{Forward $q(x_{t}|x_{t-1})$ and backward $p_{\theta}(x_{t-1}|x_{t})$ diffusion processes. $X_{0}$ is the original image, where $X_{T}$ is the Gaussian noise when $T\rightarrow \infty$.}
		\label{fig:fig2}
	\end{figure}

DDPMs were first introduced by Sohl-Dickstein \cite{sohldickstein2015deep} and then popularized by Ho et al \cite{ho2020denoising}, as a parametrized Markov chain with variational inference. Markov process defines a stochastic process where the future only depends on the present. Dhariwal and Nichol \cite{dhariwal2021diffusion} showed that DDPMs can achieve superior image quality and beat GANs.
%	\begin{equation}
%		P(X_t|X_{t-1},X_{t-2},..X_0) = P(X_t|X_{t-1})
%	\end{equation}
The diffusion process in DDPMs is inspired by the thermodynamic process of diffusion. DDPM has forward and backward diffusion processes. The forward process is defined as adding Gaussian noise to an image and the backward process is trying to obtain the original image by using the noisy one (Figure \ref{fig:fig2}).
	
For the image generation, the backward process is started from pure Gaussian noise at $t=T$ and repeated until we obtain $X_{0}$ at $t=0$.

\subsection{Conditional Generation from Unconditional Diffusion Model}

The sequential nature of the diffusion models enables researchers to interfere with the generation process. There are several studies based on the principle of altering the generation process to utilize an unconditional model to generate conditional images. In general, a measured image is defined as a noisy version of the real image transformed by a degradation operation.  % Figure \ref{fig:conditional_generation} illustrates the scheme for conditional generation using the unconditional diffusion model.
	\begin{equation}
	y = Ax+n
	\end{equation}
Degradation operation ($A$) is selected to mimic the relation of measured value ($y$) and real value ($x$) and ($n$) represents the measurement noise. The degradation depends on the inverse problem. For super-resolution, the operation will be downsampling with Gaussian blurring. For inpainting, the transformation is the multiplication of a mask operator. 

%\begin{figure}[h]
%	\centering
%	\includegraphics[width=.48\textwidth]{condition_scheme.png}
%	\caption{The scheme for conditional generation using unconditional diffusion model. Starting from the pure noise ($t=T$),  backward diffusion step generates intermediary image ($X'_{t-1}$). Then, measurement value ($y_{n}$) is used for conditioning the image onto desired manifold ($X_{t-1}$). This process is repeated until ($t=0$).}
%	\label{fig:conditional_generation}
%\end{figure}

Choi et al.\cite{choi2021ilvr} defined a process called Iterative Latent Variable Refinement (ILVR). Proposed algorithm is to generate $x'_{t-1}$ from $x_{t}$ using unconditional generation, then update predicted $x'_{t-1}$ using the condition:
	\begin{equation}
	x_{t-1} = x'_{t-1} + A^{T}(y_{t-1})- AA^{T}(x'_{t-1})
	\end{equation}
where $A^{T}$ is the reverse operation to degradation, e.g. upsampling for super-resolution. This process can be deemed as a projection operation onto the desired image manifold. The measurement is considered noiseless ($n=0$) in this setting.

%\begin{figure}[h]
%	\centering
%	\includegraphics[width=.48\textwidth]{projection.png}
%	\caption{.}
%	\label{fig:projection}
%\end{figure}

Chung et al. (a)\cite{chung2022improving} proposed an improvement to projection-based approaches named Manifold Constraint Gradient (MCG) correction. They devised a series of constraints to ensure that the gradient of the measurement remains on the manifold.

Projection-based approaches may perform well in some cases, but they have downsides. Since the predicted image is projected onto the measurement, noise in the measurement is amplified at each iteration. Chung et al. (b)\cite{chung2023diffusion} developed an alternative approach that uses gradient-based correction to the predicted image, named Diffusion Posterior Sampling (DPS). In this way, the noise in the measurement is handled during the gradient process, unlike projection-based methods.
	\begin{equation}
	x_{t-1} = x'_{t-1} - \eta\nabla_{x_{t-1}}||y-A(x'_{0})||^{2}_{2}
	\end{equation}
$\eta$ is the scaling factor for DPS, which determines the strength of the measurement in the final image. Chung et al. (b)\cite{chung2023diffusion} found that $\eta$ should be between 0.1 and 1, experimentally. Smaller $\eta$ values lead to hallucination in the generated images, while larger values may produce artifacts.

%\begin{figure}[h]
%	\centering
%	\includegraphics[width=.48\textwidth]{dps_conditioning.png}
%	\caption{.}
%	\label{fig:dps}
%\end{figure}

\subsection{Come-Closer-Diffuse-Faster}

Although the pre-trained diffusion models are robust and domain-adaptable, they are computationally expensive. Producing even a single image is a time-consuming process. Chung et al. (c)\cite{chung2022comecloserdiffusefaster} claim that for the inverse problems, starting the generation from pure Gaussian noise is redundant. Using an estimate of initial image $x_{0}$, the reverse diffusion process can start at a timestep  $t<<T$. They referred to this process as Come-Closer-Diffuse-Faster. An initial estimate can be as simple as an interpolation, or as fine-detailed as the output of a fully-trained neural network. As the initial estimate improves, the diffusion process can be shortened. This allows us to make use of existing deep learning models and go beyond their performance.
%
%\begin{figure}[h]
%	\centering
%	\includegraphics[width=.48\textwidth]{comecloser.png}
%	\caption{Come-closer-diffuse-faster algorithm.}
%	\label{fig:ccdf}
%\end{figure}

\section{Method}
\subsection{Dataset}

For all experiments, we have employed OpenAI’s unconditional DDPM model \cite{guideddiffusion} trained with ImageNet 256x256 dataset\cite{deng2009imagenet}.  Since we are using a pre-trained diffusion model, training and validation of the diffusion models are beyond the scope of this study. 

For testing the proposed method in this study, the Duke PAM \cite{anthony_dispirito_iii_2020_4042171} dataset is used. Duke PAM  is an open-source database acquired by an optical resolution PAM system at a wavelength of 532 nm in the Photoacoustic Imaging Lab at Duke University. This PAM system has a lateral resolution of 5 $\mu$m and an axial resolution of 15 $\mu$m \cite{DiSpirito2021-dl}. This dataset consists largely of in vivo images of mouse brain microvasculature, from which 18 images are randomly selected as the test set.

Given that the selected diffusion model is trained with 256x256 images, central cropping is applied to make input image dimensions 256x256. Then, images are normalized before being passed to the diffusion model. These normalized images are considered to be the ground truth images ($X_{ref}$).

At this step, our study is divided into three paths to generate input images ($y$):
\begin{enumerate}
	\item For super-resolution, $1/4$ downsampling  operation is performed and followed by low-pass filtering to prevent anti-aliasing using the Resizer Python package developed by Shocher et al.\cite{Resizer}. Therefore, the resulting image has 6.25\% of the pixels of the original image.
	\item For inpainting with a uniform undersampling, every four rows out of five is replaced with a zero, leaving 20\% of the original pixels. 
	\item For inpainting with a random undersampling, 80\% of the pixels are replaced with zero, leaving 20\% of the original pixels.
\end{enumerate}

\subsection{Neural Network as an Approximate Solution}

The previous works mainly focused on reconstructing uniformly undersampled PAM images\cite{DiSpirito2021-dl, VU2021100266}. They approached it as an inpainting problem and used U-Net shaped neural networks for a solution. In this study, we demonstrated that existing solutions can be benefitted and improved using our algorithm. To this end, we trained a U-Net model, which was subsequently subjected to comparative evaluation and involved in our proposed algorithmic framework. 

\begin{figure}[h]
	\centering
	\includegraphics[width=.48\textwidth]{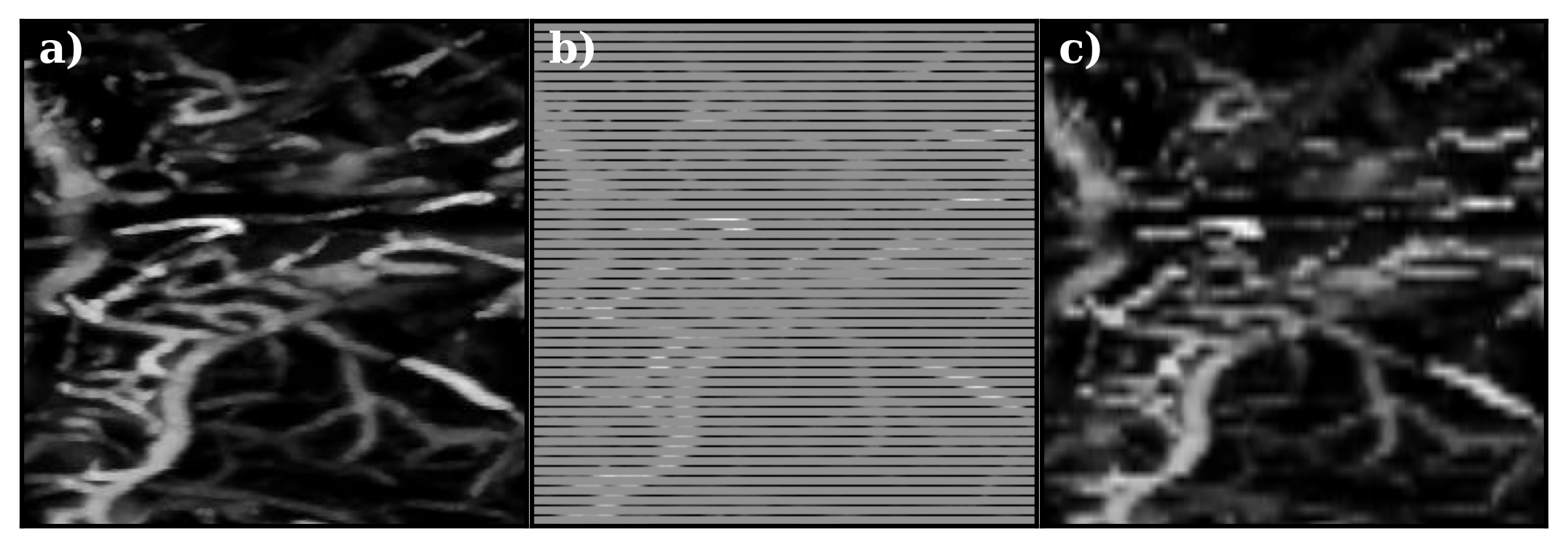}
	\caption{An example mice brain microvasculature image from Duke PAM dataset. a) Original (ground truth) image, b) uniformly undersampled (every four rows out of five is replaced with a zero) and c) missing pixels are filled via bilinear interpolation.}
	\label{fig:periodic_input}
\end{figure}

For the training, we used the 337 training samples in the Duke PAM dataset. For regularization, the data augmentation processes, namely, random crops, horizontal and vertical flips with 30\% probability, random rotations up to 20 degrees, and random Gaussian blur are applied to each image in the training set. Then images are normalized and used as the ground truth. Input images are constructed by applying a uniform undersampling to the ground truth image, which replaces every four rows out of five with a zero, leaving 20\% of the original pixels. Then, empty rows are filled with bilinear interpolation. Figure \ref{fig:periodic_input} illustrates an example image with uniform masking and the resulting bilinear interpolation.

The network was optimized using Adam optimizer\cite{kingma2017adam} of Pytorch optimizer package and $l_{1}$ loss function. We used the PyTorch framework version 1.11. The hyperparameter tuning is achieved by splitting 20\% of the training data as the validation set.

\subsection{The Reconstruction Algorithm}

%\begin{figure}[h]
%	\centering
%	\includegraphics[width=.48\textwidth]{reconstruction_algo.png}
%	\caption{.}
%	\label{fig:algorithm}
%\end{figure}

The steps of the algorithm are the following:
\begin{itemize}
	\item The operation ($A$) transforms the real value ($X$) into the measured value ($y$). For super-resolution, the operation will be downsampling. For inpainting, the transformation is the multiplication of a mask operator.
	$$ y = AX_{ref} $$
	\item Optional: An approximate solution to the inverse problem is required. The better we approximate the real value, the fewer diffusion steps would be needed. The solution depends on the transform operator $A$. But in general, interpolation operations or a simple trained neural network as an approximator can be utilized. Depending on the approximate solution, the starting point ($N<T$) for the reverse diffusion process can be settled.
	\item After having $X_{N}$, an unconditional diffusion model is employed to generate $X_{N-1}$. Then, using DPS conditioning the image generation process is iterated in the direction of interest.
\end{itemize}
Table \ref{table:experiments} summarizes the experiments conducted in this study, regarding the selection of the measurement operations ($A$), approximate solutions and the starting point for the diffusion processes ($N$).

\begin{table*}[]
	\begin{tabular}{lllll}
		\toprule
		\textbf{Inverse Problem} & \textbf{Measurement Operation}   & \textbf{Approximate Solution} & \textbf{Starting Point (N)} &  \\
		Inpainting (uniform)    & Uniform Undersampling                 & -                             & T                           &  \\
		Inpainting (uniform)    & Uniform Undersampling                 & Bilinear Interpolation        & T/2                         &  \\
		Inpainting (uniform)    & Uniform Undersampling                   & Pre-trained U-Net             & T/5                         &  \\
		Inpainting (Random)  & Random Undersampling                   & -                             & T                           &  \\
		Super Resolution       & Downsampling + Gaussian Blurring & -                             & T                           & 
	\end{tabular}
	\caption{The experimens conducted in this study, regarding the selection of the measurement operations ($A$), approximate solutions and the starting point for the diffusion processes ($N$).}
	\label{table:experiments}
\end{table*}

The diffusion model that we used in this study is trained with $T=1000$ diffusion steps. Depending on the quality of the approximate solution, we have selected $N=500$ for bilinear and bicubic interpolation and $N=200$ for pre-trained U-Net outputs.

Image production time is highly correlated with the sample step size. Selecting $N=T/k$  reduces the computing time by a factor of $k$.

\subsection{Evaluation}

The quality of reconstructed images is evaluated using three key metrics: PSNR, structural similarity index (SSIM), and learned perceptual image patch similarity (LPIPS). PSNR is the logarithm of the ratio between the peak signal value and the mean squared error (MSE) between the reference and the reconstructed images. The PSNR goes to infinity when MSE is equal to zero. On the other hand, SSIM is designed to simulate any image distortion using a combination of three factors: loss of correlation, luminance distortion, and contrast distortion\cite{5596999}. The SSIM values range between zero and one. Zero SSIM means no correlation between the images, and one refers to identical images. While PSNR and SSIM are widely acknowledged, they may occasionally fall short in aligning with human perceptual judgments. In response to this, LPIPS, as developed by Richard Zhang et al.\cite{zhang2018perceptual}, leverages deep neural networks to offer a more refined measure of image similarity that is closer to human perception. Contrary to PSNR and SSIM, the lower LPIPS scores mean better image quality. In this study, the Scikit-image framework is employed to calculate PSNR and SSIM metrics, whereas the lpips Python package with AlexNet is used to compute LPIPS scores.

\section{Results}
\label{section:results}

\subsection{Inpainting}

For inpainting, 18 images are randomly selected from the test set, and subjected to uniform and random masking leaving 20\% effective pixels.  Then, images are reconstructed using pre-trained U-Net and diffusion model conditioned on undersampled image (DiffPam). Four DiffPam experiments are conducted for inpainting, as mentioned in the Methods section. 

Producing a single 256x256 image from the $T=1000$ sample steps took 18 min on average on AWS Sagemaker Studio Lab accelerated computing. Starting from the halfway, image reconstruction time can be reduced to 9 min. Even further, starting from the U-Net model output, the time can be reduced up to 3.6 min. 

\begin{figure}[h]
	\centering
	\includegraphics[width=.48\textwidth]{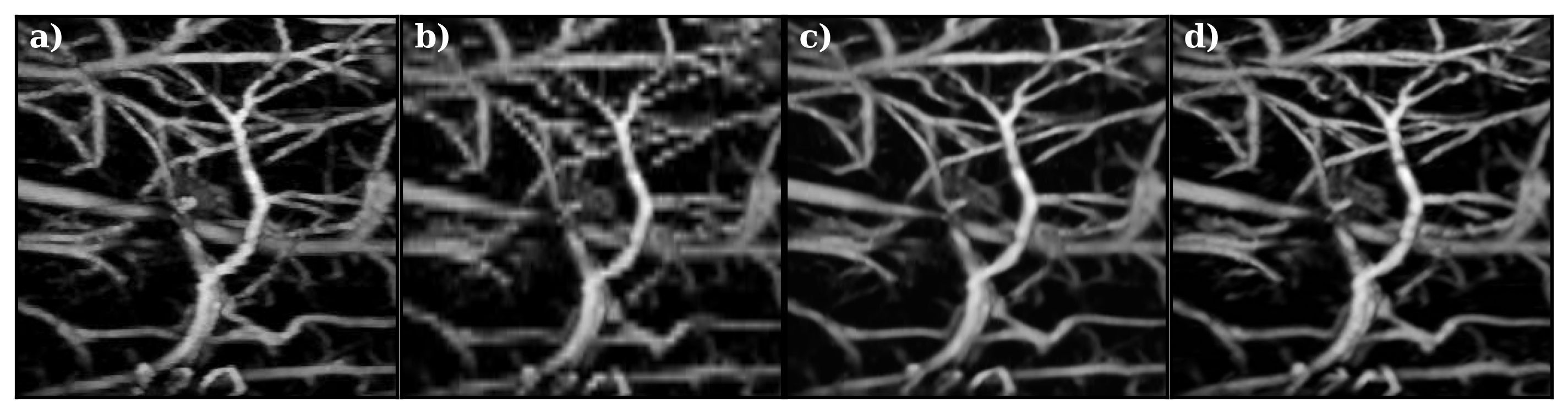}
	\caption{An example PAM image from our test set. a) Original (ground truth) image, b) bilinear interpolation applied to the uniformly undersampled image c) reconstructed image by pre-trained U-Net d) reconstructed image by DiffPam using (b) as the input.}
	\label{fig:periodic_example}
\end{figure}

Three out of four inpainting experiments were conducted using uniform masking, with differing approximate solutions. Figure \ref{fig:periodic_example} demonstrates the ground truth (fully-sampled) PAM image, bilinear interpolation, and the results of reconstruction from the undersampled version. For simplicity, only the diffusion model started with bilinear input (B-DiffPam) is illustrated in the figure.

\begin{figure*}[]
	\centering
	\includegraphics[width=0.75\textwidth]{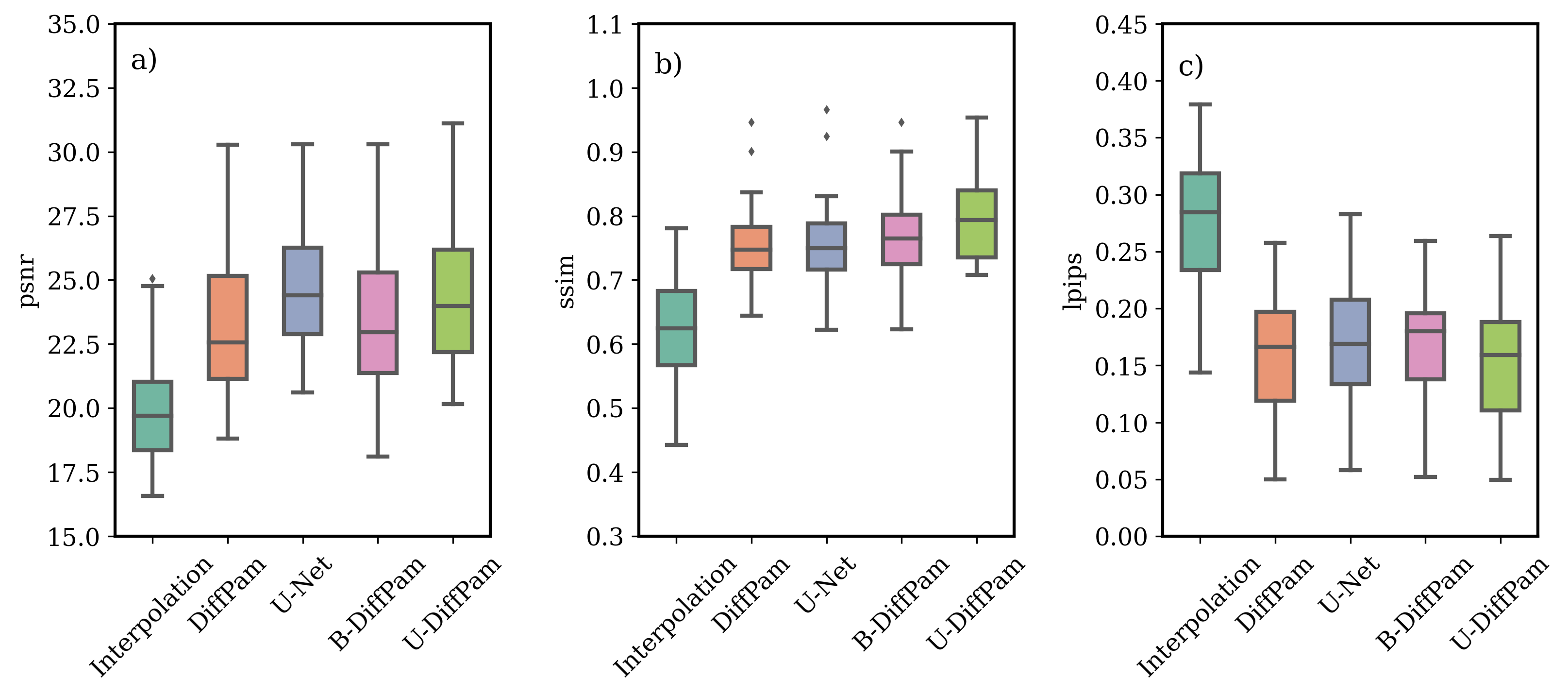}
	\caption{The uniform undersampling experimental results in terms of a) PSNR, b) SSIM and c) LPIPS metrics. DiffPam refers to the reconstruction of uniformly undersampled image using the diffusion model from pure noise. B-DiffPam refers to the diffusion process starting from the halfway with bilinear as input, lastly U-DiffPam refers to the diffusion process using the output of U-Net model.}
	\label{fig:periodic_results}
\end{figure*}

The result statistics in Figure \ref{fig:periodic_results} illustrate that, in all metrics, there is no statistical significance between the domain-specifically trained U-Net and DiffPam performances. In addition, DiffPam slightly improved the median SSIM, when applied to U-Net outputs (U-DiffPam).  

% pre-trained U-Net is superior in terms of PSNR due to being trained with this objective. Nonetheless, the U-Net tends to produce smoother outputs.

As an alternative to uniform masking, we offer scanning randomly distributed pixels with an equal number of scanned pixels. We have reconstructed the selected test images with random masking using  $T=1000$ sample steps in the same diffusion model and compared the results with uniform masking. Table \ref{table:results} demonstrates the PSNR, SSIM, and LPIPS metrics of undersampled and reconstructed images with two masking methods. As seen in the table, there are no significant differences between the metrics of the undersampled images. Nonetheless, when we reconstructed images, the results of random masking outperformed the ones of uniform masking in all metrics.

\begin{table}[]
	\begin{tabular}{@{}p{0.21\linewidth}p{0.18\linewidth}p{0.18\linewidth}p{0.185\linewidth}l}
		\toprule
		\textbf{}                              & \textbf{PSNR} & \textbf{SSIM} & \textbf{LPIPS} \\ 
		\midrule
		Undersampled (uniform)   & 7.699±2.962   & 0.096±0.062   & 0.854±0.078    \\
		Undersampled (random)  & 7.760±2.973   & 0.102±0.063   & 1.223±0.103    \\
		Reconstructed (uniform)   & 23.250±3.261  & 0.758±0.079   & 0.160±0.056    \\
		Reconstructed (random)   & 28.395±3.714  & 0.881±0.047   & 0.094±0.021   
	\end{tabular}
	\caption{The metric comparison between inputs and outputs of different masking strategies.}
	\label{table:results}
\end{table}

\begin{figure*}[]
	\centering
	\includegraphics[width=.8\textwidth]{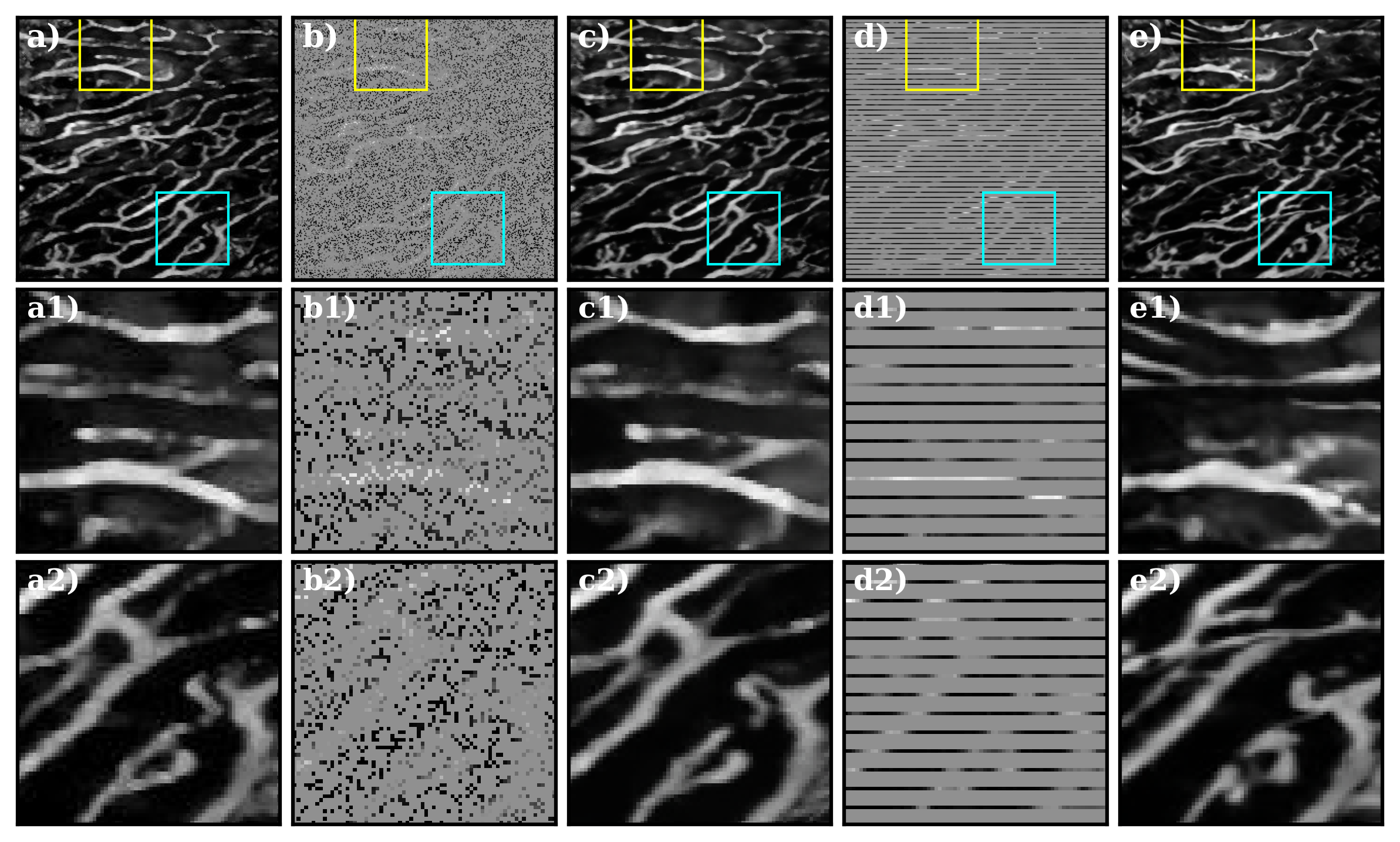}
	\caption{Visual comparison of random and uniform inpainting DiffPam results. a) Original (ground truth) image b) undersampled image with random masking c) reconstructed image from the random mask d) undersampled image with uniform masking e) reconstructed image from the uniform mask. The second row illustrates the focus of the upper box, whereas the third row demonstrates the focus of the lower box. e1) and e2) contain artifacts in the microvasculature, where c1) and c2) resembles the ground truth.}
	\label{fig:inpainting_compare}
\end{figure*}

Upon examining the reconstructed images, we noted the presence of certain artifacts in microvasculature exclusively in uniformly masked images, a phenomenon absent in their randomly masked counterparts (Figure \ref{fig:inpainting_compare}). Although we are aware that acquiring signals from random locations in PAM is a more challenging task than orienting row by row, the benefits of achieving it may surpass its hardship.

\subsection{Super Resolution}

\begin{figure*}[]
	\centering
	\includegraphics[width=.8\textwidth]{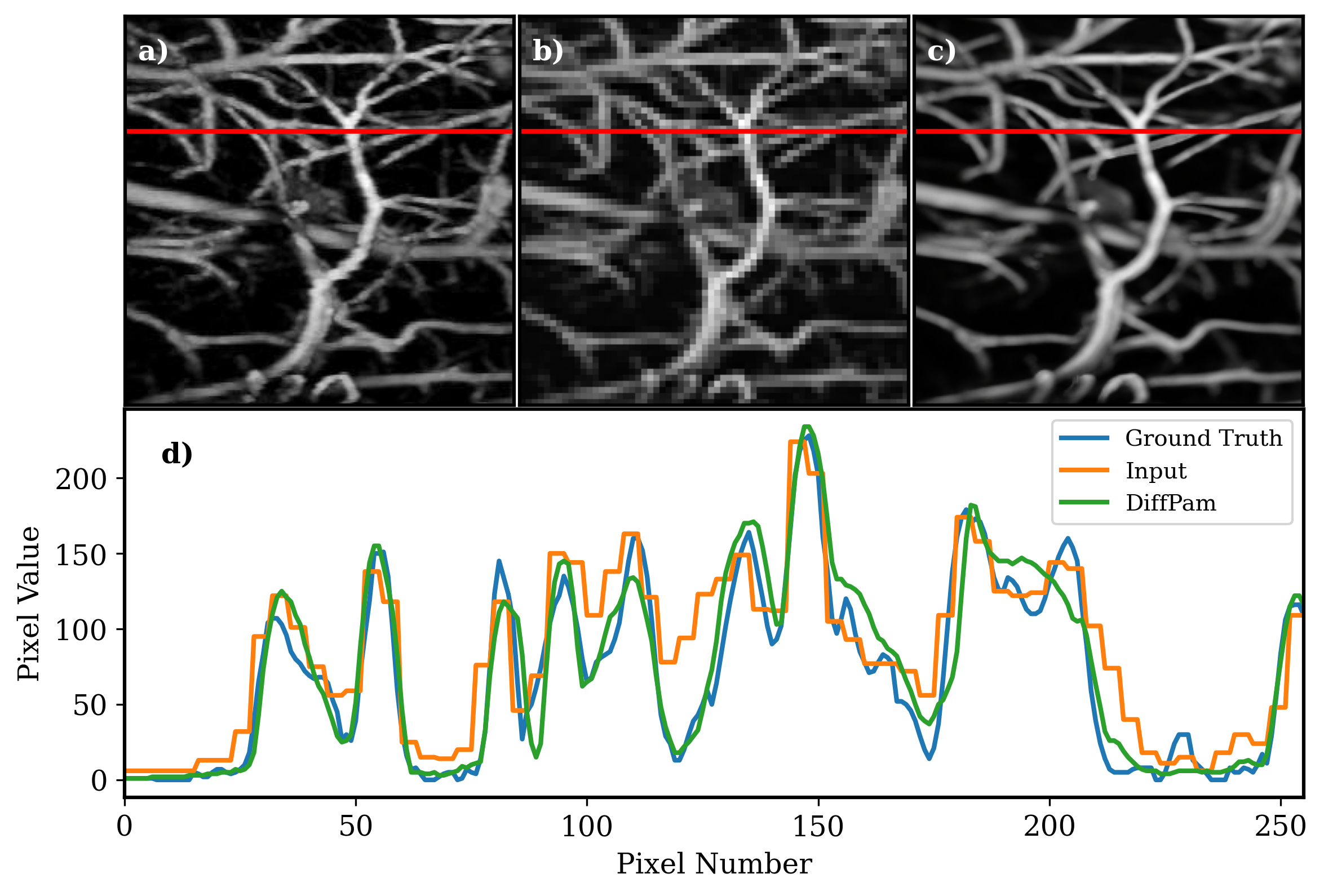}
	\caption{The example result of super-resolution from 4x downsampled image. a) Original (ground truth) image, b) 4x downsampled image, c) reconstructed image by DiffPam, d) pixel values across the red line in the images. DiffPam is able to reconstruct the downsampled images using prior knowledge of natural images. }
	\label{fig:superres}
\end{figure*}

If acquiring signals in random positions is deemed an impossible task, we suggest increasing the spatial solution. The corresponding operation might be a downsampling with Gaussian blurring in the image processing domain. We experimented with this setting using six images in the test set. We were able to improve the PSNR by 18.2\% and SSIM by 31.7\%, and lower the LPIPS by 54.0\% on average. Figure \ref{fig:superres} illustrates the ground truth, downsampled and reconstructed images, and pixel value change across the red line. 

% To supplementary materials
% If we compare the the results of diffusion starting from scratch (DiffPam) and from the bilinear interpolation input (B-DiffPam), we observed that DiffPam outperforms the PSNR metric, but gives similar results in terms of SSIM and LPIPS.

\section{Discussion}
\label{section:discussion}

In this study, we demonstrated that the proposed algorithm DiffPam, adeptly reconstructs undersampled PAM images without the need of a large dataset or training a deep learning model. The diffusion model which is solely trained with the ImageNet database of natural images, was able to perform on par with the U-Net model specifically trained with mice brain microvasculature image database. Furthermore, for users with an existing trained model, our algorithm provides a means to enhance results without interfering with the original model.

As the days go by, the advancing artificial intelligence technology constantly demands more computing power; however, accessing high computing power remains a challenge among scientists, especially with limited resources. We value the removal of these obstacles in the progress of science. Hence, we demonstrated that the diffusion process can be shortened to reduce the amount of computing and processing time, with limited or zero accuracy loss.

\begin{figure}[]
	\centering
	\includegraphics[width=.45\textwidth]{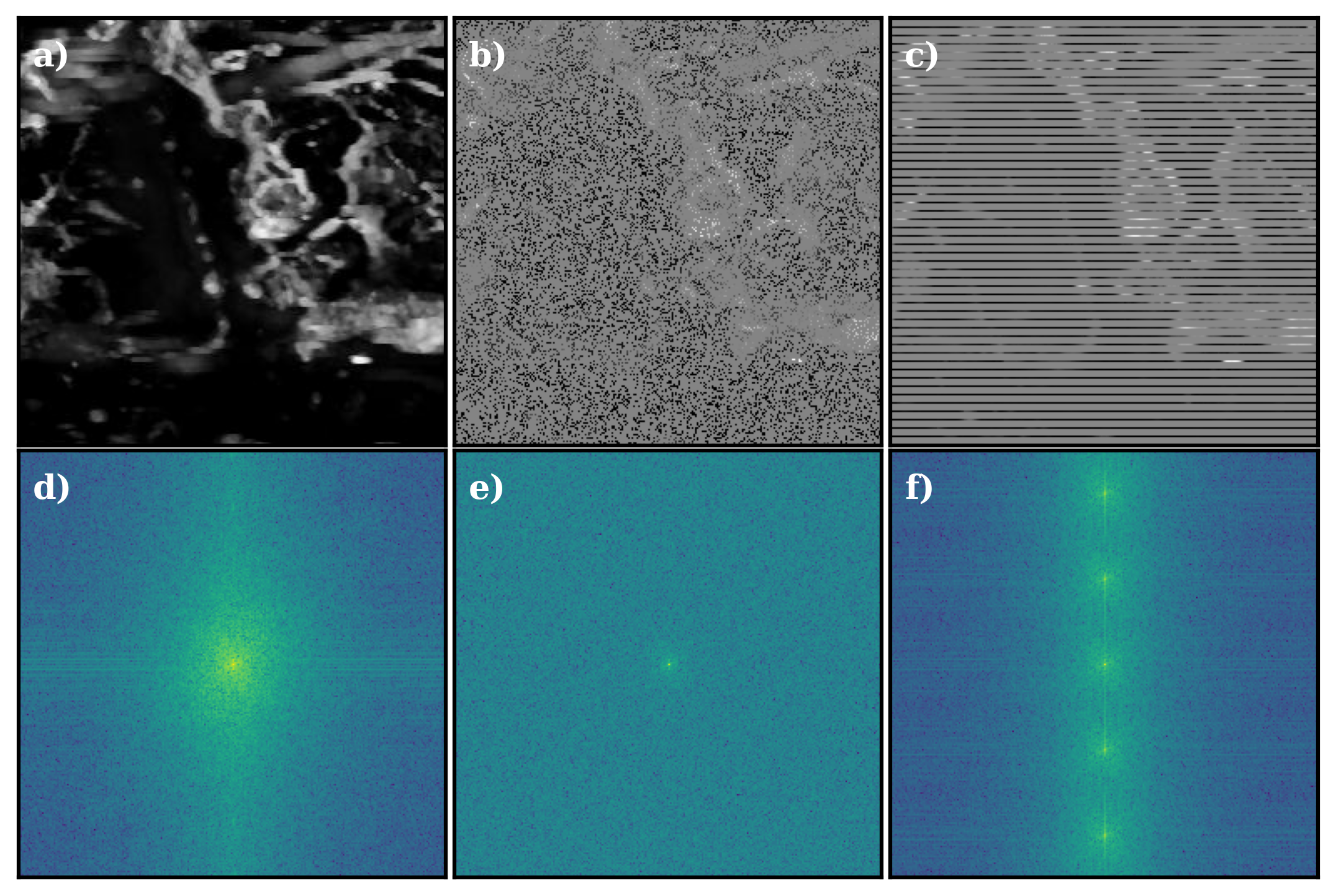}
	\caption{Visual comparison of a) original, b) randomly masked, and c) uniformly masked microvasculature photoacoustic images and their Fourier components (d-f). The Fourier components of the uniformly masked image (c) present unmatched coefficients compared to the original image (d).}
	\label{fig:fourier}
\end{figure}

Since the speed of the scanning axes of PAM imaging may differ greatly, skipping every few lines in the slower axis is a more convenient way of speeding up the process of acquiring images. uniform masking of PAM images, which is commonly used in the literature, is a way of representing this process. Nonetheless, removing rows or columns periodically from an image, results in disappearing components in the frequency (Fourier) domain (Figure \ref{fig:fourier}). Thus, some valuable information is lost in the process, making the lossless recovery impossible. The information loss is the primary cause of the artifacts in reconstructed images, which are illustrated in the section \ref{section:results}. These artifacts may cause misinterpretation of medical data, which can be hazardous. Hence, we highlight that adapting random scanning of pixels generates more accurate and reliable results than uniform sampling. An alternative approach can be employing larger spatial resolution, and then using a super-resolution approach to reconstruct finer details. In this study, we aimed to demonstrate the limitations of all approaches and leave the decision to the researchers.

% Therefore, adapting random masking with the same number of pixels scanned generates superior results than periodical masking. An alternative approach can be employing larger spatial resolution, then using a super-resolution approach to reconstruct finer details. In this study, we aimed to demonstrate the limitations of all approaches and leave the decision to the researchers.

% Another crucial aspect is that generative models tend to hallucinate, i.e. create images which are not faithful to original ones.  

The primary result of our study is the potential utilization of the proposed algorithm by researchers with limited expertise or computational resources in artificial intelligence. The diffusion model employed in our investigation is openly accessible and can be substituted with any other proficient diffusion model. Our source code is accessible on a public GitHub repository\footnote{https://github.com/iremzog/diffpam}. For scientists facing constraints in accessing accelerated computing, the option of fast-tracked, truncated diffusion steps presents a viable solution for reducing computation time.

Our study is limited to a single image dataset of narrow scope, namely mice brain microvasculature. Experiments are made with a relatively small number of images, due to limited access to accelerated computing resources. Nonetheless, we hope that our study will shed light on further research in this field.

\section{Conclusion}

In conclusion, our study introduces the DiffPam algorithm, showcasing its efficiency in reconstructing undersampled PAM images without the need for extensive datasets or deep learning model training. The diffusion model, exclusively trained on the natural image database, demonstrated comparable performance to an in-domain trained neural network. We also address the escalating demand for computing power in advancing AI technology, emphasizing the significance of overcoming obstacles in accessing high computational resources for scientific progress. We propose the reduction of computing time through shortened diffusion processes without compromising accuracy. Additionally, our exploration of random or uniform sampling techniques in PAM imaging underscores the superiority of random sampling in preserving valuable information. We acknowledge the limitations of our study, confined to a specific image dataset, and a relatively small sample size due to resource constraints. However, we anticipate that our findings will inspire further research in this domain, offering researchers with limited resources a valuable algorithmic tool for PAM image reconstruction. Our algorithm's accessibility is underscored by the availability of source code on a public GitHub repository, providing a practical way for researchers to implement and extend our work.

% \section*{acknowledgements}

\section*{conflict of interest}
The authors declare no conflict of interest.

\bibliography{references}

\printendnotes

\end{document}

\section{Supplementary Material}
\beginsupplement

\subsection{Denoising Diffusion Probabilistic Models}

Forward diffusion is a Markov chain process is defined as 
\begin{equation}
	q(x_{t}|x_{t-1}) = \mathcal{N}(\sqrt{1-\beta_{t}} x_{t-1},\,\beta_{t}I)
\end{equation}
, where $\beta_{0}<\beta_{1}<…<\beta_{T}$. The $\beta_{t}$’s are defined as the noise scheduler. Reparameterizing the $\alpha_{t} := 1-\beta_{t}$$ \alpha_{t} := 1-\beta_{t}$$\bar{\alpha_{t}} := \Pi^{t}_{i=1} \alpha_{i}$, we can calculate $x_{t}$ from $x_{0}$ at arbitrary time t:
\begin{equation}
	\begin{split}
		q(x_{t}|x_{0}) & = \mathcal{N}(\sqrt{\bar{\alpha_{t}}}x_{0}, (1-\bar{\alpha_{t}})I) \\
		x_{t} & = \sqrt{\bar\alpha_{t}} x_{0} + \epsilon \sqrt{1-\bar\alpha_{t}}
	\end{split}
\end{equation}
For the reverse process;
\begin{equation}
	p_{\theta}(x_{t-1}|x_{t}) = \mathcal{N}(x_{t-1},\hat\mu_{\theta}(x_{t},t),\Sigma_{\theta}(x_{t},t))
\end{equation}
A denoiser deep learning model can be used to perform the backward diffusion process. There are three equivalent objective functions to optimize the model: 
\begin{itemize}
	\item predicting the original image  $x_{0}$,
	\item predicting the noise  $\epsilon$,
	\item predicting the score function  $\nabla \log p(x_{t})$.
\end{itemize}

Predicting the original image $x_{0}$ is equivalent to learning to predict the noise; empirically, however, Ho et al. have found that predicting the noise resulted in better performance\cite{ho2020denoising}.
	\begin{equation}
			\hat\mu_{\theta}(x_{t},t) = \frac{1}{\sqrt{\alpha_{t}}}(x_{t} - \frac{\beta_{t}}{\sqrt{1-\bar\alpha_{t}}}\epsilon_{\theta}(x_{t},t))
	\end{equation}

Ho et al. \cite{ho2020denoising} proposed to use fixed variance due to the hardship of model optimization to predict the variance. They found that variance can be fixed to $\beta_{t}$ or $\tilde\beta_{t} := \beta_{t} \frac{1-\bar\alpha_{t-1}}{1-\bar\alpha_{t}}$ without changing the performance. Since $\beta_{t}$ and $\tilde\beta_{t}$ are very close except at the very beginning of the diffusion process, the result may seem reasonable. However, the beginning of the diffusion process makes a crucial contribution. Therefore, Dhariwal and Nichol et al.\cite{dhariwal2021diffusion} proposed learning the variance which is a linear combination of $\beta_{t}$ and $\tilde\beta_{t}$ in the log domain. 

\begin{equation}
	\Sigma_{\theta}(x_{t},t) = exp(v.\log\beta_{t} + (1-v).\log\tilde\beta_{t})
\end{equation}

In order to optimize over this new variance, they proposed hybrid objective that combines the loss function of the noise $\epsilon$ and the variance ($	\Sigma_{\theta}(x_{t},t) $). In this study, we utilized a diffusion model trained with the hybrid objective. 
	\begin{equation}
			L_{hybrid} = L_{simple} + \lambda L_{vlb}
		\end{equation}

%	
%\begin{figure}[h]
%	\centering
%	\includegraphics[width=.48\textwidth]{backward_scheme.png}
%	\caption{Final scheme for the backward calculation.}
%	\label{fig:fig4}
%\end{figure}

\subsection{Compressed Sensing}

In the signal acquiring process, the sampling step should be smaller than half of the desired spatial resolution according to the Nyquist theorem. Undersampling below the Nyquist rate creates the artiacts named aliasing in the reconstructed signal. Compressed sensing (CS) is an optimization technique enables sampling much lower than Nyquist sampling rate and reconstructing the signal using non-linear algorithm \cite{1614066}. Compressed sensing has a wide application in magnetic resonance imaging (MRI) and computed tomography(CT) \cite{4472246, https://doi.org/10.1118/1.2836423}. This technique has two prerequisite:
\begin{itemize}
	\item The signal must be sparce and
	\item The undersampling method must be incoherent in the given basis.
\end{itemize}
Under these conditions, CS suggests that a signal can be perfectly reconstructed using the undersampled version using $L_{1}$ minimization. Our method is similar with compressed sensing in the sense that we are minimizing the $L_{1}$ loss between measured (undersampled) and generated image. Therefore, the incoherent undersampling method (random masking) produces favourable results than coherent (uniform) bases. 
 
\subsection{More Results}

\begin{figure*}[]
	\centering
	\includegraphics[width=.9\textwidth]{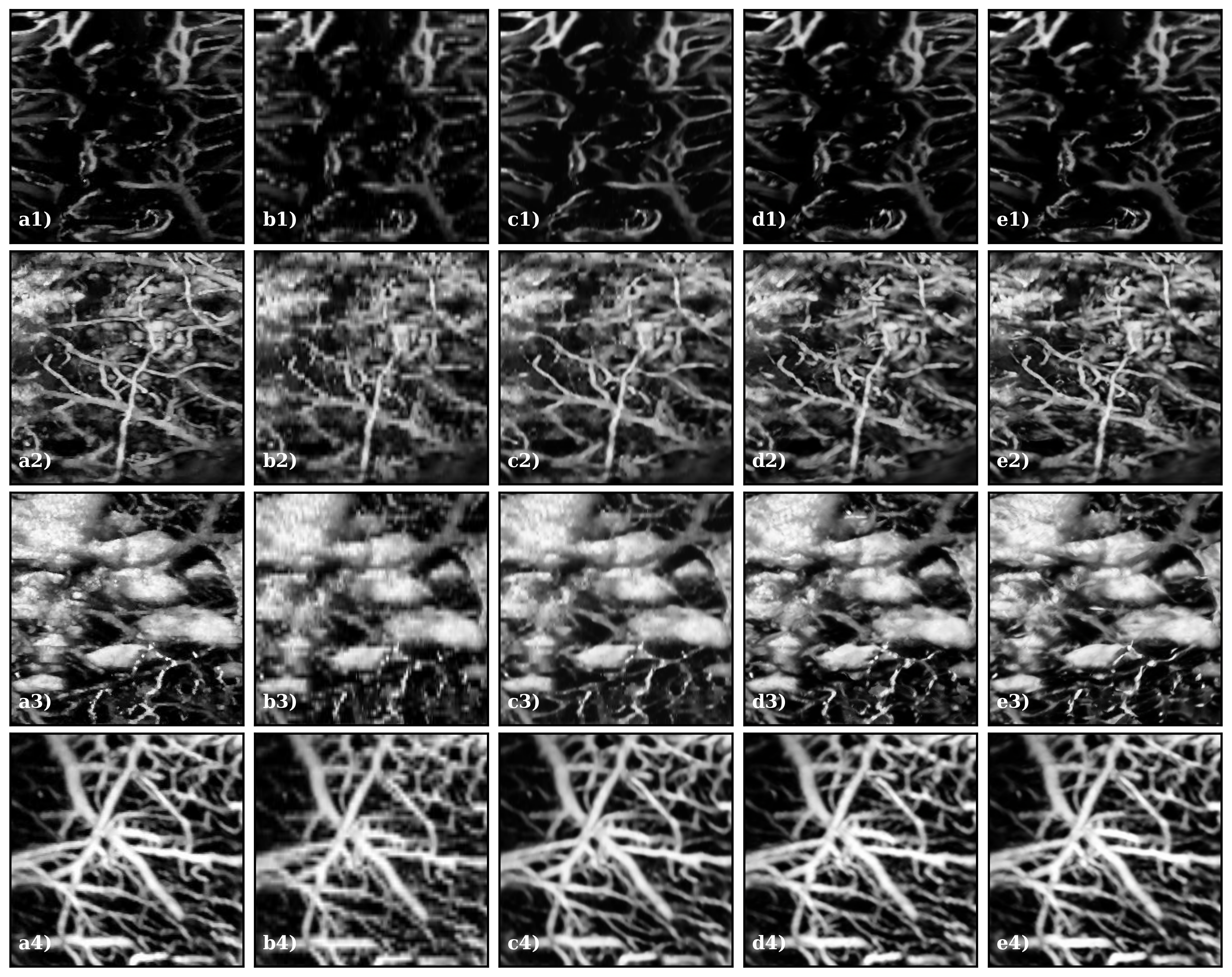}
	\caption{Visual comparison of U-Net and DiffPam results. a) Original (ground truth) image b) bilinear interpolation applied to the uniformly undersampled image c) reconstructed image by pre-trained U-Net d) reconstructed image by DiffPam with bilinear input e) reconstructed image by DiffPam with noise input.}
	\label{fig:periodic_more_results}
\end{figure*}

\begin{figure*}[]
	\centering
	\includegraphics[width=.9\textwidth]{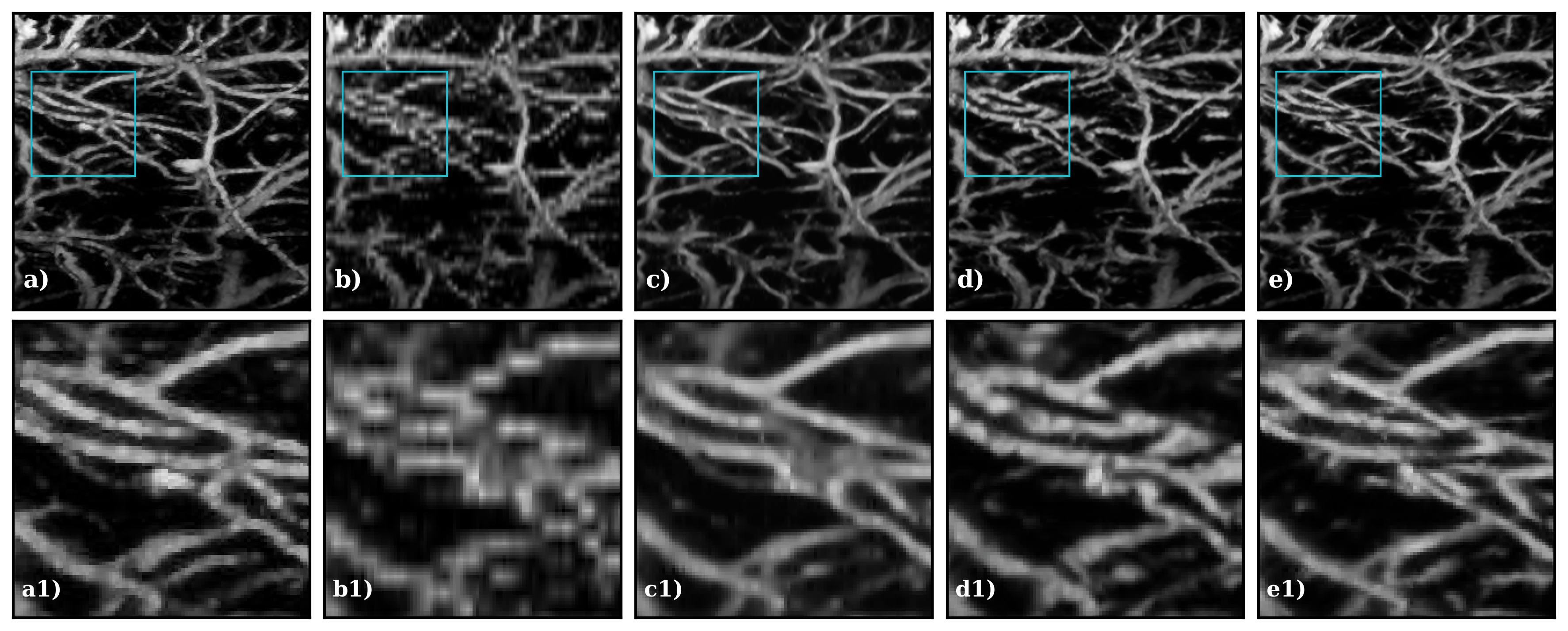}
	\caption{Visual comparison of U-Net and DiffPam results. a) Original (ground truth) image b) bilinear interpolation applied to the uniformly undersampled image c) reconstructed image by pre-trained U-Net d) reconstructed image by DiffPam with bilinear input e) reconstructed image by DiffPam with noise input. The second row illustrates the focus of the box. All reconstuction methods display some level of artifacts in the microvasculature structure.}
	\label{fig:periodic_fail}
\end{figure*}

\begin{figure*}[]
	\centering
	\includegraphics[width=.9\textwidth]{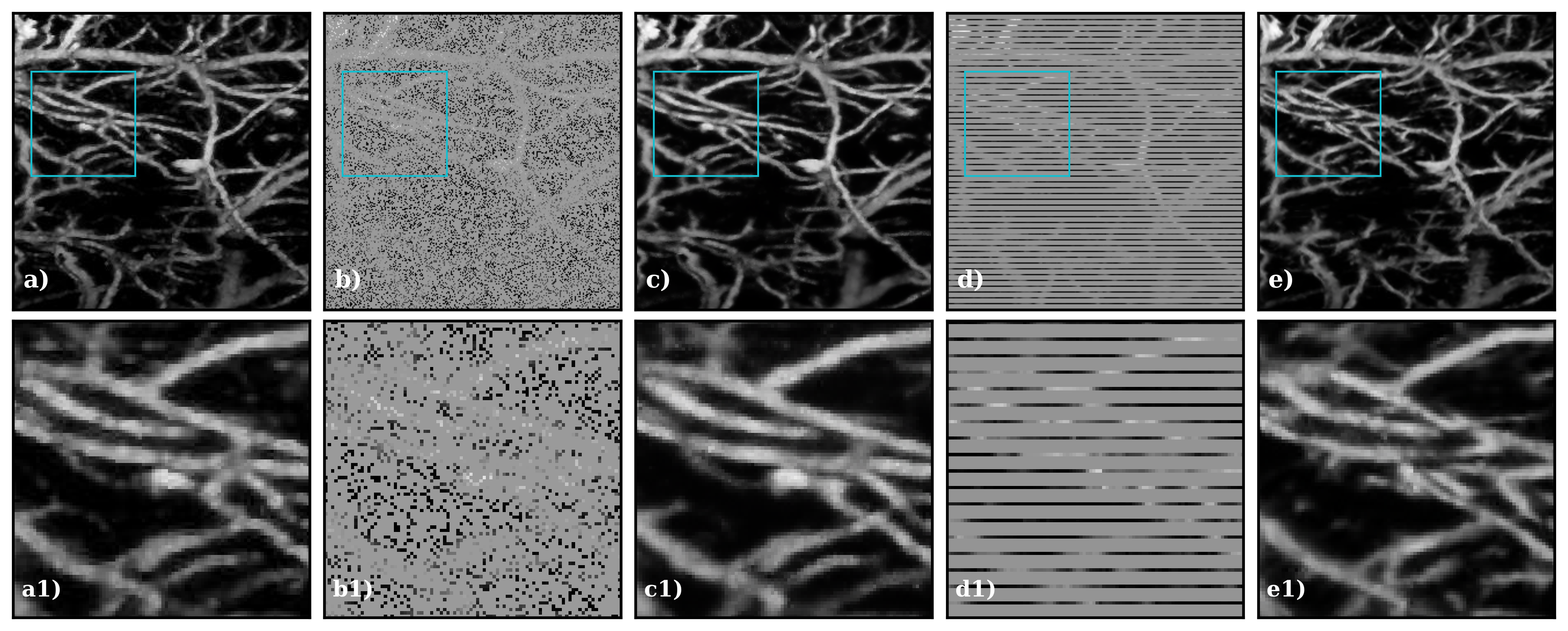}
	\caption{Visual comparison of random and uniform inpainting DiffPam results. a) Original (ground truth) image b) undersampled image with random masking c) reconstructed image from the random mask d) undersampled image with uniform masking e) reconstructed image from the uniform mask. The second row illustrates the focus of the box. e1) contain artifacts in the microvasculature, where c1) resembles the ground truth}
	\label{fig:periodic_fail2}
\end{figure*}

\FloatBarrier

\bibliography{references}

\printendnotes